\documentclass[11pt,a4paper]{article}

\usepackage{comment}

\usepackage[paper=a4paper,margin=3.9cm]{geometry}

\usepackage[backend=bibtex,
firstinits,style=numeric-comp,maxnames=99]{biblatex} %

\bibliography{shortnames.bib, bib-latest.bib}

\usepackage{graphics}
\usepackage[noend]{algpseudocode}

\usepackage{amsmath, amsfonts, amsthm}
\usepackage{amssymb}
\usepackage{xspace}
\usepackage[T1]{fontenc}
\usepackage[normalem]{ulem}
\usepackage{braket}
\usepackage{tikz}
\usetikzlibrary{calc}

\theoremstyle{theorem}
\newtheorem{theorem}{Theorem}[]

\newtheorem{lemma}[theorem]{Lemma}
\newtheorem{corollary}[theorem]{Corollary}
\renewcommand{\geq}{\geqslant}
\renewcommand{\leq}{\leqslant}
\renewcommand{\epsilon}{\varepsilon}

\newcommand{\calA}{\ensuremath{\mathcal{A}}\xspace}
\newcommand{\calB}{\ensuremath{\mathcal{B}}\xspace}

\newcommand{\calL}{\ensuremath{\mathcal{L}}\xspace}
\newcommand{\calM}{\ensuremath{\mathcal{M}}\xspace}
\newcommand{\calT}{\ensuremath{\mathcal{T}}\xspace}

\newcommand{\defeq}{\stackrel{\scriptscriptstyle \text{def}}{=}}

\newcommand{\ileft}{\ensuremath{i_\text{left}}\xspace}
\newcommand{\iright}{\ensuremath{i_\text{right}}\xspace}

\newcommand{\SigmaT}{\ensuremath{\Sigma_{\textup{T}}}}
\newcommand{\SigmaP}{\ensuremath{\Sigma_{\textup{P}}}}
\newcommand{\Sm}{\ensuremath{S}}                        %
\newcommand{\SmFilt}{\ensuremath{S_{\textup{filt}}}}    %
\newcommand{\SigmaFilt}{\ensuremath{\Sigma_{\textup{filt}}}}
\newcommand{\SigmaLast}{\ensuremath{\Sigma_{\textup{last}}}}

\newcommand{\upto}{\ensuremath{,\,\,}} %
\newcommand{\pred}[1]{\ensuremath{\textup{pred\textbf{(}}#1{\textup{\textbf{)}}}}}
\newcommand{\phid}[1]{\ensuremath{\phi \text{\big(} #1 \text{\big)}}}

\newcommand{\fplong}[1]{\ensuremath{\Phi_{#1}}\xspace}

\newcommand{\fpinner}[1]{\ensuremath{\Phi^0_{#1}}\xspace}
\newcommand{\fpdiff}{\ensuremath{\Delta}}

\newcommand{\pmatch}{\mbox{p-match}\xspace}
\newcommand{\pmatches}{\mbox{p-matches}\xspace}
\newcommand{\pperiod}{\mbox{p-period}\xspace}

\newcommand{\prooflemma}[1]{\textbf{Proof of Lemma~\ref{#1}}}
\newcommand{\prooftheorem}[1]{\textbf{Proof of Theorem~\ref{#1}}}

\newcommand{\fsplit}{\ensuremath{\ominus}}

\newcommand{\fzero}{\ensuremath{\circledcirc}}

\newcommand{\procA}{\textup{A}\xspace}
\newcommand{\procB}{\textup{B}\xspace}
\newcommand{\procBdelta}{\textup{B1}\xspace}
\newcommand{\procBphi}{\textup{B2}\xspace}
\newcommand{\procC}{\textup{C}\xspace}

\newcommand{\insertdiagram}[1]{
    \centering
    \includegraphics[scale=0.70]{./diagrams/#1}}

\newcommand{\insertfigure}[2]{
    \begin{figure}[t]
        \centering
        \insertdiagram{#1}
        \caption{#2}
    \end{figure}
    }

\title{Parameterized Matching in the Streaming Model} %

\author{
    Markus Jalsenius\thanks{~Department of Computer Science, University of Bristol,  U.K.} \and
    Benny Porat\thanks{~Department of Computer Science, Bar-Ilan University, Israel.}\and
    Benjamin Sach\thanks{~Department of Computer Science, University of Warwick, U.K.}}

\date{}

\begin{document}

\maketitle

\begin{abstract}

We study the problem of parameterized matching in a stream where we want to output matches between a pattern of length $m$ and the last $m$ symbols of the stream before the next symbol arrives. Parameterized matching is a natural generalisation of exact matching where an arbitrary one-to-one relabelling of pattern symbols is allowed. We show how this problem can be solved in constant time per arriving stream symbol and sublinear, near optimal space with high probability. Our results are surprising and important: it has been shown that almost no streaming pattern matching problems can be solved (not even randomised) in less than $\Theta(m)$ space, with exact matching as the only known problem to have a sublinear, near optimal space solution. Here we demonstrate that a similar sublinear, near optimal space solution is achievable for an even more challenging problem. The proof is considerably more complex than that for exact matching.
 \end{abstract}

\section{Introduction}

We consider the problem of pattern matching in a stream where we want to output matches between a pattern of length $m$ and the last $m$ symbols of the stream. Each answer must be reported before the next symbol arrives. The problem we consider in this paper is known as \emph{parameterized matching} and is a natural generalisation of exact matching where an arbitrary one-to-one relabelling of the pattern symbols is allowed (one per alignment). For example, if the pattern is \texttt{abbca} then there there is a parameterized match with \texttt{bddcb} as we can apply the relabelling \texttt{a}$\rightarrow$\texttt{b}, \texttt{b}$\rightarrow$\texttt{d}, \texttt{c}$\rightarrow$\texttt{c}. There is however no parameterized match with \texttt{bddbb}. We show how this streaming pattern matching problem can be solved in near constant time per arriving stream symbol and sublinear, near optimal, space with high probability. The space used is reduced even further when only a small subset of the symbols are allowed to be relabelled.  As discussed in the next section, our results demonstrate a serious push forward in understanding what pattern matching algorithms can be solved in sublinear space.

\subsection{Background}

Streaming algorithms is a well studied area and specifically finding patterns in a stream is a fundamental problem that has received increasing attention over the past few years. It was shown in~\cite{CEPP:2008} that many offline algorithms can be made online (streaming) and deamortised with a $\log m$ factor overhead in the time complexity per arriving symbol in the stream, where $m$ is the length of the pattern. There have also been improvements for specific pattern matching problems but they all have one property in common: space usage is $\Theta(m)$ words. It is not difficult to show that we in fact \emph{need} as much as $\Theta(m)$ space to do pattern matching, unless errors are allowed.
The field of pattern matching in a stream took a significant step forwards in 2009 when it was shown to be possible to solve exact matching using only $O(\log{m})$ words of space and $O(\log{m})$ time per new stream symbol~\cite{Porat:09}.  This method, which is based on fingerprints, correctly finds all matches with high probability. The initial approach was subsequently somewhat simplified~\cite{EJS:2010} and then finally improved to run in constant time~\cite{BG:2011} within the same space requirements.

Being able to do exact matching in sublinear space raised the question of what other streaming pattern matching problems can be solved in small space. In~2011 this question was answered for a large set of such problems~\cite{CJPS:2011}. The result was rather gloomy: almost no streaming pattern matching problems can be solved in sublinear space, not even using randomised algorithms. An $\Omega(m)$ space lower bound was given for $L_1$, $L_2$, $L_\infty$, Hamming, edit distance and pattern matching with wildcards as well as for any algorithm that computes the cross-correlation/convolution. So what other pattern matching problems could possibly be solved in small space? It seems that the only hope to find any is by imposing various restrictions on the problem definition. This was indeed done in~\cite{Porat:09} where a solution to $k$-mismatch (exact matching where up to $k$ mismatches are allowed) was given which uses $O(k^2\text{poly}(\log m))$ time per arriving stream symbol and $O(k^3\text{poly}(\log m))$ words of space. The solution involves multiple instances of the exact matching algorithm run in parallel. Note that the space bound approaches $\Theta(m)$ as $k$ increases, so the algorithm is only interesting for sufficiently small $k$. Further, the space bound is very far from the known $\Omega(k)$ lower bound.
We also note that it is straightforward to show that exact matching with $k$~wildcards in the pattern can be solved with the $k$-mismatch algorithm. To our knowledge, no other streaming pattern matching have been solved in sublinear space so far.

In this paper we present the first push forward since exact matching by giving a sublinear, near optimal space and near constant time (or constant with a mild restriction on the alphabet) algorithm for parameterized matching in a stream. This natural problem turns out to be significantly more complicated to solve than exact matching and our results provide the first demonstration that small space and time bounds are achievable for a more challenging problem. Note that our space bound, as opposed to $k$-mismatch, is essentially optimal like for exact matching.
One could easily argue that our results are surprising, and yet again the question of what other problems are solvable in sublinear space calls for an answer. In particular, given that restrictions to the problem have to be made, what restrictions should one make to break the $\Omega(m)$ space barrier.

\subsection{Problem definition and related work}\label{sec:probdef}

A pattern $P$ of length $m$ is said to \emph{parameterize match}, or \emph{\pmatch} for short, an $m$~length string $S$ if there is an injective (one-to-one) function $f$ such that $S[j]=f(P[j])$ for all $j\in\{0,\dots,m-1\}$.
In our streaming setting, the pattern is known in advance and
the symbols of the stream $T$ arrive one at a time.  We use the letter $i$ to denote the index of the latest symbol in the stream. Our task is to output whether there is a \pmatch between $P$ and $T[(i-m+1)\upto i]$ before $T[i+1]$ arrives. The mapping $f$ may be distinct for each $i$.

One may view this matching problem as that of finding matches in a stream encrypted using a substitution cipher. In offline settings, parameterized matching has its origin in finding duplication and plagiarism in software code although has since found numerous
other applications.  Since the first introduction of the problem, a great deal of work has gone into
its study in both theoretical and practical settings (see
e.g.\@~\cite{Baker:1993,AFM:1994,
  Baker:1995,Baker:1996,Baker:1997,HLS:2007}). Notably, in an offline
setting, the exact parameterized matching problem can be solved in
near linear time using a variant~\cite{AFM:1994} of the classic linear
time exact matching algorithm KMP~\cite{Knuth:1977}.

When the sublinear space algorithm for exact matching was given in~\cite{Porat:09},
properties of the periods of strings formed a crucial part of their analysis. However, when considering parameterized matching the period of a string is a much less straightforward concept than it is for exact matching.
For example, it is no longer true that consecutive matches must either be separated by the period of the pattern or be at least $m/2$ symbols apart. This property, which holds for exact but not parameterized matching, allows for an efficient encoding of the positions of the matches. This was crucial to reducing the space requirements of the previous streaming algorithms.  Unfortunately, parameterized matches can occur at arbitrary positions in the stream, requiring new insights. This is not the only challenge that we face.

A natural way to match two strings under parameterization is to consider
their \emph{predecessor strings}. For a string $S$, the predecessor
string, denoted $\pred{S}$, is a string of length $|S|$ such that
$\pred{S}[j]$ is the distance, counted in numbers of symbols, to the
previous occurrence of the symbol $S[j]$ in $S$. In other words,
$\pred{S}[j]=d$, where $d$ is the smallest positive value for
which $S[j]=S[j-d]$. Whenever no such $d$ exists, we set
$\pred{S}[j]=0$.
As an example, if $S=\texttt{aababcca}$ then $\pred{S}=\texttt{01022014}$.
We can perform parameterized matching offline by only considering predecessor strings using the fundamental fact~\cite{Baker:1993} that two equal length strings $S$ and $S'$ \pmatch iff $\pred{S}=\pred{S'}$.
A plausible approach for our streaming problem would now be to
translate the problem of parameterized matching in a stream to that of
exact matching.  This could be achieved by converting both pattern and
stream into their corresponding predecessor strings and maintaining
fingerprints of a sliding window of the translated input. However,
consider the effect on the predecessor string, and hence its
fingerprint, of sliding a window in the stream along by one.  The
leftmost symbol $x$, say, will move out of the window and so the
predecessor value of the new leftmost occurrence of $x$ in the new
window will need to be set to \texttt{0} and the corresponding fingerprint
updated. We cannot afford to store the positions of all
characters in a $\Theta(m)$ length window. %

We will show a matching algorithm that solves these problems and others we
encounter en route using minimal space and in near constant time per arriving symbol. A number of technical innovations are required, including new uses of fingerprinting, a new compressed encoding of the positions of potential matches, a separate deterministic algorithm designed  for prefixes of the pattern with small parameterized period as well as the deamortisation of the entire matching process. Section~\ref{sec:overview} gives a more detailed overview of these main hurdles.

\subsection{Our new results}\label{sec:newresults}

Our main result is a fast and space efficient algorithm to solve the streaming parameterized matching problem.
It applies to \emph{dense} alphabets where we assume that both the pattern and streaming text alphabets are $\Sigma=\{0,\dots,|\Sigma|-1\}$. The following theorem is proved over the subsequent sections of this paper.

\begin{theorem}\label{thm:main}
    Suppose the pattern and text alphabets are both $\Sigma=\{0,\dots,|\Sigma|-1\}$ and the pattern has length~$m$. There is a randomised algorithm for streaming parameterized matching that takes $O(1)$ worst-case time per character and uses $O(|\Sigma|\log m)$ words of space. The probability that the algorithm outputs correctly at all alignments of an $n$~length text is at least $1-1/n^c$, where $c$ is any constant.
\end{theorem}

To fully appreciate this theorem we also give a nearly matching space lower bound which shows that our solution is optimal within logarithmic factors. The proof is based on communication complexity arguments and is deferred to Appendix~\ref{appendix:space}.

\begin{theorem}
\label{thm:space}
    There is a randomised space lower bound of  $\Omega(|\Sigma|)$ bits for the streaming parameterized problem, where $\Sigma$ is the pattern alphabet.
\end{theorem}

Parameterized matching is often specified under the assumption that only some symbols are variable (allowed to be relabelled). The mapping $f$ we used in Section~\ref{sec:probdef} has to reflect this constraint. More precisely, let the pattern alphabet be partitioned into fixed symbols $\Sigma_{\textup{fixed}}$ and variable symbols $\Pi$. For $\sigma\in\Sigma_{\textup{fixed}}$, we require that $f(\sigma)=\sigma$. The result from Theorem~\ref{thm:main} can be extended to handle \emph{general} alphabets with arbitrary fixed symbols. The idea is to apply a suitable reduction that was given in~\cite{AFM:1994} (Lemma~2.2) together with the streaming exact matching algorithm of Breslauer and Galil~\cite{BG:2011}, as well as applying a ``filter'' on the text stream, using for instance the
the dictionary of Andersson and Thorup~\cite{AT:2000} based on exponential search trees. The dictionary is used to map text symbols to the variable pattern symbols in $\Pi$. The proof of the following theorem is given in Appendix~\ref{appendix:filter}.

\begin{theorem}
    \label{thm:filter}
    Suppose $\Pi$ is the set of pattern symbols that can be relabelled under parameterized matching. All other pattern symbols are fixed. Without any constraints on the text alphabet, there is a randomised algorithm for streaming parameterized matching that takes $O(\sqrt{\log{|\Pi|}/\log{\log{|\Pi|}}})$ worst-case time per character and uses $O(|\Pi|\log m)$ words of space, where $m$ is the length of the pattern. The probability that the algorithm outputs correctly at all alignments of an $n$ length text is at least $1-1/n^c$, where $c$ is any constant.
\end{theorem}

As part of the proof of Theorem~\ref{thm:main} we had to develop an algorithm that efficiently solves streaming parameterized matching for patterns with small
\emph{parameterized period}, defined as follows. The parameterized period (\emph{\pperiod}) of the pattern $P$, denoted $\rho$, is the smallest positive integer such that $P[0\upto (m-1-\rho)]$ \pmatches $P[\rho\upto m-1]$. That is, $\rho$ is the shortest distance that $P$ must be slid by to parameterized match itself.
Our algorithm is deterministic and is interesting in its own right (see Section~\ref{sec:smallrho} for details). We also provide a matching space lower bound which is detailed in Appendix~\ref{appendix:space}.

\begin{theorem}
    \label{thm:kmp-main}
    Suppose the pattern and text alphabets are both $\Sigma=\{0,\dots,|\Sigma|-1\}$ and the pattern has \pperiod~$\rho$. There is a deterministic algorithm for streaming parameterized matching that takes $O(1)$ worst-case time per character and uses $O(|\Sigma|+\rho)$ words of space. Further, there is a deterministic space lower bound of $\Omega(|\Sigma| + \rho)$ bits.
\end{theorem}

\subsection{Fingerprints}
We will make extensive use Rabin-Karp style fingerprints of strings
which are defined as follows. Let $S$ be a string over the alphabet $\Sigma$. Let $p > |\Sigma|$ be a prime and choose
$r \in \mathbb{Z}_p$ uniformly at random.  The
fingerprint $\phi(S)$ is given by $\phi(S) \defeq \sum_{k=0}^{|S|-1} S[k] r^k \bmod p$. A critical  property of the fingerprint function $\phi$ is that the probability of achieving a false positive, $\Pr(\phi(S) = \phi(S') \,\land\, S \ne S')$, is at most $|S|/(p-1)$ (see~\cite{KR:1987, Porat:09} for proofs). Let $n$ denote the total length of the stream. Our randomised algorithm will make $o(n^2)$ (in fact near linear) fingerprint comparisons in total. Therefore, by the applying the union bound, for any constant $c$, we can choose $p \in \Theta(n^{c+3})$ so that with probability at least $1-1/n^c$ there will be no false positive matches.

As we assume the RAM model with word size $\Theta(\log{n})$, a fingerprint fits in a constant number of words. We assume that all fingerprint arithmetic is performed within $\mathbb{Z}_p$.
In particular we will take advantage of two fingerprint operations.
\begin{itemize}
\item[\fsplit]   \emph{Splitting:} Given $\phid{S[0 \upto a]}$, $\phid{S[0 \upto b]}$ (where $b>a$) and the value of $r^{-a} \bmod p$, we can compute $\phid{S[a+1 \upto b]} = \phid{S[0 \upto b]} \fsplit \phid{S[0 \upto a]}$ in $O(1)$ time.%

\item[\fzero] \emph{Zeroing:} Let $S, S'$ be two equal length strings such that $S'$ is identical to $S$ except for in positions $z\in Z \subseteq [0,s-1]$ at which $S'[z]=0$. We write $\phid{S} \fzero Z$ to denote $\phid{S'}$. Given $\phid{S}$ and $(S[z],\,r^z \bmod p)$ for all $z\in Z$, computing $\phid{S} \fzero Z$ takes $O(|Z|)$ time.

\end{itemize}

\section{Overview, key properties and notation}\label{sec:overview}

The overall idea of our algorithm in Theorem~\ref{thm:main} follows that of previous work on streaming exact matching in small space, however for parameterized matching the situation is much more complex and calls for not only more involved details and methods but also a deep fundamental understanding of the nature of parameterized matching. We will now describe the overall idea, introduce some important notation and at the end of this section we will highlight  key facts about parameterized matching that are crucial for our solution. %

The main algorithm will try to match the streaming text with various prefixes of the pattern~$P$. Let $\SigmaP$ denote the pattern alphabet.
We define $\delta = |\SigmaP|\log{m}$ and
let $P_0$ denote the shortest prefix of $P$ that has
\pperiod greater than $3\delta$ (recall the definition of \pperiod given above Theorem~\ref{thm:kmp-main}).
We define $s$ prefixes $P_\ell$ of increasing length so that
$|P_{\ell}| = 2^{\ell}|P_0|$ for $\ell \in \{1, \ldots, s-1\}$, where $s \leq \lceil \log m \rceil$ is the largest value such that $|P_{s-1}|\leq m/2$. The final prefix $P_s$ has length $m-4 \delta$.
For all $\ell$, we define $m_\ell = |P_\ell|$, hence $m_\ell=2m_{\ell-1}$. %

In order to determine if there is a \pmatch between the text and a pattern prefix, we will compare the fingerprints of their predecessor strings (recall that two strings \pmatch iff their predecessor strings are the same).
We will need two related (but typically distinct) fingerprint definitions to achieve this. Figure~\ref{fig:fp-output} will be helpful when reading the following definitions which are discussed in an example below.
For any index~$i'$ and~$\ell \in \{0, \ldots, s\}$,
\begin{align*}
    \fplong{\ell}(i') \,&\defeq\, \phid{\pred{T[0 \upto (i'+m_{\ell}-1)]}}\,, \\
    \fpinner{\ell}(i') \,&\defeq\, \phid{\pred{T[i'\upto (i'+m_{\ell}-1)]}[m_{\ell-1}\upto m_{\ell}-1]}\,.
\end{align*}

\insertfigure{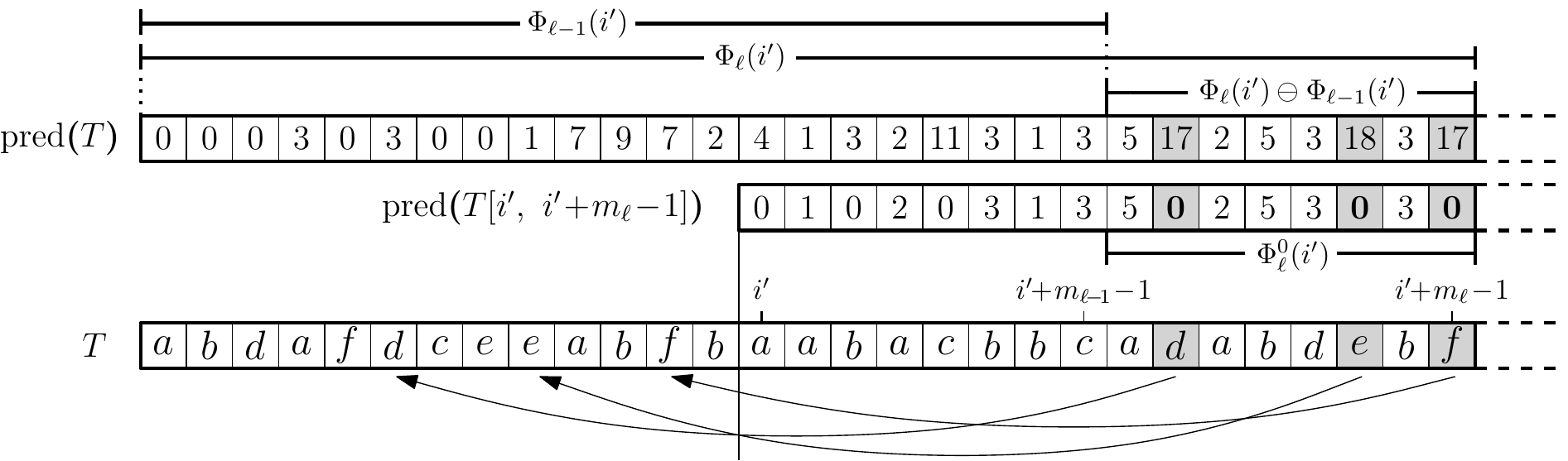}{\label{fig:fp-output} The key fingerprints used by the randomised algorithm.
Characters contribute differently to $\fpinner{\ell}(i')$ and $\fplong{\ell}(i')\ominus\fplong{\ell-1}(i')$ are highlighted.}

For each $\ell \in \{1, \ldots, s\}$ the main algorithm runs a process whose responsibility for finding \pmatches between the text and $P_\ell$ ($P_0$ is handled separately as will be discussed later). The process responsible for $P_\ell$ will ask the process responsible for $P_{\ell-1}$ if it has found any \pmatches, and if so it will try to extend the matches to $P_\ell$. As an example, suppose that the process for $P_{\ell-1}$ finds a match at position $i'$ of the text (refer to Figure~\ref{fig:fp-output}). The process will then store this match along with the fingerprint $\fplong{\ell-1}(i')$ which has been built up as new symbols arrive. The process for $P_\ell$ will be handed this information when the symbol at position $i'+m_\ell-1$ arrives. The task is now to work out if $i'$ is also a matching position with $P_\ell$. With the fingerprint $\fplong{\ell}(i')$ available (built up as new symbols arrive), the process for $P_\ell$ can use fingerprint arithmetics to determine if $i'$ is a matching position. This is one instance where the situation becomes more tricky than one might first think.

As position $i'$ is a \pmatch with $P_{\ell-1}$ it suffices to compare the second half of the predecessor string of $P_\ell$ with the second half of the predecessor string of $T[i'\upto i+m_\ell-1]$. Fingerprints are used for this comparison. It is crucial to understand that $\fplong{\ell}(i')\fsplit \fplong{\ell-1}(i')$ cannot be used directly here;
some predecessor values of the text might point very far back, namely to some position \emph{before} index $i'$. In Figure~\ref{fig:fp-output} we have shaded the three symbols for which this is true and we have drawn arrows indicating their predecessors. Thus, in order to correctly do the fingerprint comparison we need to set those positions to zero (we want the fingerprint of the predecessor string of the text substring starting at position $i'$, not the beginning of $T$). The fingerprint we defined as $\fpinner{\ell}(i')$ above is the fingerprint we want to compare to the fingerprint of the second half of the predecessor string of $P_\ell$. Using fingerprint operations, we have from the definitions that $\fpinner{\ell}(i') = \big( \fplong{\ell}(i')\fsplit \fplong{\ell-1}(i') \big) \fzero \fpdiff_\ell(i')$, where $\fpdiff_\ell(i')$ is the set of positions that have to be set to zero. For a substring of $T$ of length $\Theta(m_{\ell-1})$ consider the subset of positions which occur in $\fpdiff_\ell(i')$ for at least one value of $i'$. Any such position has a predecessor value greater than $m_{\ell-1}$. Therefore, by summing over all distinct symbols we have that the size of this subset is crucially only $O(|\SigmaP|)$. Thus, we can maintain in small space every position in a suitable length window that will \emph{ever} have to be set to zero.

Let us go back to the example where the process for $P_{\ell-1}$ had found a \pmatch at position~$i'$. The process stores $i'$ along with the fingerprint $\fplong{\ell-1}(i')$. This information is not needed by the process for $P_\ell$ until $m_{\ell-1}$ text symbols later. During the arrival of these symbols, the process for $P_{\ell-1}$ might detect more \pmatches, in fact many more matches. Their positions and corresponding fingerprints have to be stored until needed by the process for $P_\ell$. We now have a space issue: how do we store this information in small space? To appreciate this question, first consider exact matching. Here matches are known to be either an exact period length apart or very far apart. The matching positions can therefore be represented by an arithmetic progression. Further, the fingerprints associated with the matches in an arithmetic progression can easily be stored succinctly as one can work out each one of the fingerprints from the first one.
For parameterized matching the situation is much more complex: matches can occur more chaotically and, as we have seen above, fingerprints must be updated dynamically to reflect that symbols could be mapped differently in two distinct alignments. Handling these difficulties in small space (and small time complexity) is a main hurdle and is one point at which our work differ significantly from all previous work on streaming matching in small space.
We cope with this space issue in the next section.

\subsection{The structure of parameterized matches}
First recall that an \emph{arithmetic progression} is a sequence of numbers such that the (common) difference between any two successive numbers is constant. We can specify an arithmetic progression by its start number, the common difference and the length of the sequence.
In the next lemma we will see that the positions at which a string $P$ of length $m$ parameterize matches a longer string of length $3m/2$ can be stored in small memory: either a matching position belongs to an arithmetic progression or it is one of relatively few positions that can be listed explicitly in $O(|\SigmaP|)$ space.
The proof of the lemma (consult Figure~\ref{fig:typ-matches}) is deferred to Section~\ref{sec:arithmetic-proof}.

\insertfigure{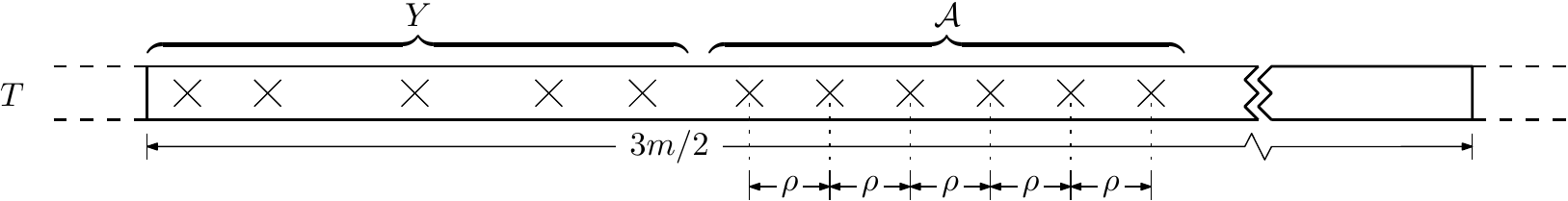}{\label{fig:typ-matches} Partitioning of positions (\!{\Large$\times$}\!) at which $P$ \pmatches in a $3m/2$ length substring of $T$.}

\begin{lemma}
    \label{lem:arithmetic}
     Let $X$ be the set of positions at which $P$ \pmatches within an $3m/2$ length substring of $T$. The set $X$ can be partitioned into two sets $Y$ and $\calA$ such that $|Y|\leq 6|\SigmaP|$, $\max(Y)<\min(\calA)$ and $\calA$ is an arithmetic progression with common difference $\rho$, where $\rho$ is the \pperiod of $P$.
\end{lemma}

The lemma is incredibly important for the algorithm as it allows us to store all partial matches (that need to be kept in memory before being discarded) in a total of $O(|\SigmaP|\log m)$ space across all processes. The question of how to store their associated fingerprints remains, but is nicely resolved with the corollary below that follows immediately from the proof of Lemma~\ref{lem:arithmetic}. We can afford to store fingerprints explicitly for the positions that are identified to belong to the set $Y$ from Lemma~\ref{lem:arithmetic}, and for the matching positions in the arithmetic progression $\calA$ we can, as for exact matching, work out every fingerprint given the first one.

\begin{corollary}
    \label{cor:arithmetic}
     For pattern $P$, text $T$ and arithmetic progression $\calA$ as specified in Lemma~\ref{lem:arithmetic}, $\pred{T}[(i+m-\rho) \upto (i+m-1)]$ is the same for all $i\in\calA$.

\end{corollary}

\subsection{Deamortisation}\label{sec:deamortisation}

So far we have described the overall approach but it is of course a major concern how to carry out computations in constant time per arriving symbol. In order to \emph{deamortise} the algorithm, we run a separate process responsible for the pattern prefix $P_0$ that uses the deterministic algorithm of Section~\ref{sec:smallrho} (i.e.~Theorem~\ref{thm:kmp-main}). As $P_0$ has $\pperiod$ greater than $3\delta$, the \pmatches it outputs are at least this far apart. This enables the other processes to operate with a small delay: process $P_\ell$ expects process $P_{\ell-1}$ to hand over matches and fingerprints with a small delay, and it will itself hand over matches and fingerprints to $P_{\ell+1}$ with a small delay. One of the reasons for the delays is that processes operate in a round-robin scheme -- one process per arriving symbol. The process that is responsible for $P_s$ (which has length $m-4\delta$) returns matches with a delay of up to $3\delta$ arriving symbols. Hence there is a gap of length $\delta$ in which we can work out if the whole of $P$ matches. To do this we have another process that runs in parallel with all other processes and explicitly checks if any match with $P_s$ can be extended with the remaining $4\delta$ symbols by directly comparing their predecessor values with the last $4\delta$ predecessor values of the pattern. This job is spread out over $\delta$ arriving symbols, hence matches with $P$ are outputted in constant time.

\section{The main algorithm}

We are now in a position to describe the full algorithm of Theorem~\ref{thm:main}. Recall that the algorithm will find \pmatches with
each of the pattern prefixes $P_0,\dots,P_s$ defined in the previous section. If a shorter prefix fails to match at a given
position then there is no need to check matches for longer
prefixes.
Our algorithm runs three main processes concurrently which we label \procA, \procB and \procC.
The term process had a slightly different meaning in the previous section, but hopefully this will cause no confusion.
Each process takes $O(1)$ time per arriving symbol. Recall that both the pattern and text alphabets are $\SigmaP=\{0,\dots,|\SigmaP|-1\}$. \textbf{Process~\procA} finds \pmatches with prefix $P_0$ which are inserted as they occur into a \emph{match queue} $M_0$. \textbf{Process \procB} finds \pmatches for prefixes $P_1,\dots,P_s$ which are inserted into the match queues $M_1, \ldots, M_s$, respectively.
The \pmatches are inserted with a delay of up to $3\delta$
symbol arrivals after they occur. \textbf{Process~\procC} finds \pmatches with the whole pattern $P$ which are outputted in constant time as they occur as described in Section~\ref{sec:deamortisation}.

It is crucial for the space usage that the match queues $M_0,M_1, \ldots, M_s$ will be stored in a compressed fashion.  The delay in detecting \pmatches with $P_\ell$ in Process \procB is a consequence of deamortising the work required to find a prefix match, which we spread out over $\Theta(\delta)$ arriving symbols. We can afford to spread out the work in this way because the \pperiod of $P_{\ell-1}$ is at least $\delta$ so any \pmatches are at least this far apart.
Throughout this section we assume that $m>14\delta$ so that $m_\ell-m_{\ell-1} \geq 3\delta$ for $\ell \in \{1, \ldots, s\}$.
If $m\leq 14\delta$, or the \pperiod of $P$ is $3\delta$ or less, we use the deterministic algorithm presented in Section~\ref{sec:smallrho} to solve the problem within the required bounds.

\subsection{Process \procA (finding matches with $P_0$)}

From the definition of $P_0$ we have that if
we remove the final character (giving the string $P[0\upto m_0-2]$) then its \pperiod is at most $3\delta$.
The \pperiod of $P_0$ itself could be much larger.
As part of process \procA we run the deterministic pattern matching algorithm from Section~\ref{sec:smallrho} (see Theorem~\ref{thm:kmp-main}) on $P[0\upto m_0-2]$. It returns \pmatches in constant time and uses $O(|\SigmaP|+3\delta)=O(|\SigmaP|\log m)$ space.

In order to establish matches with the whole of $P_0$ we handle the final character separately.
If the deterministic subroutine reports a match that ends in $T[i-1]$, when $T[i]$ arrives we have a \pmatch with $P_0$ if and only if $\pred{T}[i]=\pred{P_0}[m_0-1]$ (or $\pred{T}[i]\geq m_0$ if $\pred{P_0}[m_0-1]=0$). As the alphabet is of the form $\SigmaP=\{0,\ldots |\SigmaP|-1\}$, we can compute the value of $\pred{T}[i]$ in $O(1)$ time by maintaining an array $A$ of length $|\SigmaP|$ such that for all $\sigma \in \SigmaP$, $A[\sigma]$ gives the index of the most recent occurrence of symbol $\sigma$.

Whenever Process \procA finds a match with $P_0$ at position $i'$ of the text, the pair $(i',\fplong{0}(i'))$ is added to a (FIFO) queue $M_0$, which is queried by Process~\procB when handling prefix $P_1$.

\subsection{Process \procB (finding matches with $P_1,\dots,P_s$)}%

We split the discussion of the execution of Process~\procB into $s$ \emph{levels}, $1,\dots,s$.
For each level~$\ell$ the fingerprint $\fpinner{\ell}(i')$ is computed for each position $i'$ at which $P_{\ell-1}$ \pmatches.
Then, as discussed in Section~\ref{sec:overview}, if $\fpinner{\ell}(i') = \phi(\pred{P_\ell}[m_{\ell-1}\upto (m_\ell-1)])$, there is also a match with $P_\ell$ at $i'$. The algorithm will in this case add the pair $(i',\fplong{\ell}(i'))$ to the queue $M_\ell$ which is subject to queries by level~$\ell+1$.
To this end we compute $\fplong{\ell}(i')\fsplit\fplong{\ell-1}(i')$ and $\fpdiff_\ell(i')$, where $\fpdiff_\ell(i')$ contains all the positions which should be zeroed in order to obtain $\fpinner{\ell}(i')$.
In the example of Figure~\ref{fig:fp-output},
$\fpdiff_\ell(i')=\{1,5,7\}$ (the \texttt{d}, \texttt{e} and \texttt{f}, respectively).

In order for process \procB to spend only constant time per arriving symbol, all its work must be scheduled carefully. The preparation of the $\fpdiff_\ell(i')$ values takes place as a subprocess we name \procBdelta. Computing $\fplong{\ell}(i')\fsplit\fplong{\ell-1}(i')$ and establishing matches takes place in another subprocess named \procBphi. The two subprocesses are run in sequence for each arriving symbol.
We now give their details.

\paragraph{Subprocess \procBdelta~(prepare zeroing)}
We use a queue $D_\ell$ associated with each level~$l$ which contains the most recent $O(|\SigmaP|)$ positions with predecessor the values greater than $m_{\ell-1}$. We will see below that $\fpdiff_\ell(i')$ is a subset of the positions in $D_\ell$ (adjusted to the offset $i'$).

Unfortunately, in the worst case, for an arriving symbol $T[i]$, $i$ could belong to all of the $D_\ell$ queues. Since we can only afford constant time per arriving symbol, we cannot insert $i$ into more than a constant number of queues. The solution is to buffer arriving symbols.
When some $T[i]$ arrives we first check whether $\pred{T}[i]>m_0$. If so, the pair $(i,\pred{T}[i])$ is added to a buffer $\calB$ to be dealt with later.
Together with the pair we also store the value $r^i \bmod p$ which will be needed to perform the required zeroing operations.

In addition to adding a new element to the buffer $\calB$, the Subprocess~\procBdelta will also process elements from $\calB$. If is is currently not in the state of processing an element, it will now start doing so by removing an element from $\calB$ (unless $\calB$ is empty). Call this element $(j,\pred{T}[j])$. Over the next $s$ arriving symbols the Subprocess~\procBdelta will do the following. For each of the $s$ levels $\ell$, if $\pred{T}[j]>m_{\ell-1}$, add $(j,\pred{T}[j])$ to the queue $D_\ell$. If $D_\ell$ contains more than $12|\SigmaP|$ elements, discard the oldest.

\paragraph{Subprocess \procBphi~(establish matches)} %
This subprocess schedules the work across the levels in a round-robin fashion by only considering level $\ell = 1 + (i \bmod s)$ when the symbol $T[i]$ arrives.  Potential matches may not be reported by this subprocess until up to $3\delta$ arriving symbols after they occur. As $P_{\ell-1}$ has \pperiod at least $3\delta$, the processing of potential matches does not overlap.

The Subprocess \procBphi for level $\ell$ is always in one of two states: either it is \emph{checking} whether a matching position $i'$ for $P_{\ell-1}$ is also a match with $P_\ell$, or it is \emph{idle}. If idle, level~$\ell$ looks into queue $M_{\ell-1}$ which holds matches with $P_{\ell-1}$. If $M_{\ell-1}$ is
non-empty, level~$\ell$ removes an element from $M_{\ell-1}$, call this element $(i',\fplong{\ell-1}(i'))$, and enters the checking state. Whenever $i>i'+m_\ell+\delta$,
level~$\ell$ will start checking if $i'$ is also a matching position with $P_\ell$. It does so by first computing the fingerprint $\fplong{\ell}(i')\fsplit \fplong{\ell-1}(i')$,
which by definition equals $\big(\fplong{\ell}(i')-\fplong{\ell-1}(i')\big) r^{-i'-m_{\ell-1}} \bmod p$. We can ensure the fingerprint $\fplong{\ell}(i')$ is always available when needed by maintaining a circular buffer of the most recent $\Theta(\delta)$ fingerprints of the text. Similarly we can obtain $r^{-i'-m_{\ell-1}}\bmod p$ in $O(1)$ time by keeping a buffer of the most recent $\Theta(\delta)$ values of $r^{-i}\bmod p$ along with $r^{-m_{\ell}}\bmod p$ for all $\ell$. \label{page:buff}

Over the next at most $|\SigmaP|$ arriving symbols for which Subprocess~\procBphi is considering level~$\ell$ (i.e. those with $\ell = 1 + (i \bmod s)$),
$\fpinner{\ell}(i')$ will be computed from
$\fplong{\ell}(i') \fsplit \fplong{\ell-1}(i')$  by stepping through the elements of the queue $D_\ell$. For any element $(j,\pred{T}[j]) \in D_\ell$, we have that $(j-i'-m_{\ell-1}) \in \fpdiff_\ell(i')$ if and only if $\pred{T}[j]>j-i'$.
Further, as Subprocess~\procBdelta stored $r^j \bmod p$ with the element in $D_\ell$ and $r^{i'} \bmod p$ is obtained through the circular buffer as above, we can perform the zeroing in $O(1)$ time.

 Having computed $\fpinner{\ell}(i')$, we then compare it to $\phi(\pred{P_\ell}[m_{\ell-1}\upto (m_\ell-1)])$. If they are equal, we have a \pmatch with $P_\ell$ at position $i'$ of the text, and the pair $(i',\fplong{\ell}(i'))$ is added to the queue $M_\ell$. This occurs before $T[i'+m_\ell+3\delta]$ arrives.

\subsection{Correctness, time and space analysis}

The time and space complexity almost follow immediately from the description of our algorithm, but a little more attention is required to verify that the algorithm actually works. In particular one has to show that buffers do not overflow, elements in queues are dealt with before being discarded and every possible match will be found (disregarding the probabilistic error in the fingerprint comparisons). The proof of the next lemma is given in Appendix~\ref{appendix:correctness}.

\begin{lemma}
    \label{lem:correctness}
    The algorithm described above proves Theorem~\ref{thm:main}.
\end{lemma}

\section{The deterministic matching algorithm}\label{sec:smallrho}

We now describe the deterministic algorithm that solves Theorem~\ref{thm:kmp-main}. Its running time is $O(1)$ time per character and it uses $O(|\SigmaP|+\rho)$ words of space, where $\rho$ is the parameterized period
of $P$. We require that both the pattern and text alphabets are $\SigmaP=\{0,\dots,|\SigmaP|-1\}$.

We first briefly summarise the overall approach of \cite{AFM:1994} which our algorithm follows. It resembles the classic KMP algorithm. When $T[i]$ arrives, the overall goal is to calculate the largest $r$ such that $P[0 \upto r-1]$ \pmatches $T[(i- r+1)\upto i]$.
A \pmatch occurs iff $r=m$.
When a new text character $T[i+1]$ arrives the algorithm compares
$\pred{P}[r]$ to $\pred{T}[i+1]$ in $O(1)$ time to determine whether $P[0 \upto r]$ \pmatches $T[(i- r+1)\upto i+1]$.
    More precisely, the algorithm checks whether either $\pred{P}[r]=\pred{T}[i+1]$, or $\pred{P}[r]=0 \,\wedge\, \pred{T}[i+1] > r$. The second case covers the possibility that the previous occurrence in the text was outside the window.
If there is a match, we set $r \leftarrow r +1$ and $i \leftarrow i+1$ and continue with the next text character. If not, we shift the pattern prefix $P[0 \upto r-1]$ along by its \pperiod, denoted $\rho_{r-1}$, so that it is aligned with $T[(i- r+\rho_{r-1}+1)\upto i]$. This is the next candidate for a \pmatch. In the original algorithm, the p-periods of all prefixes are stored in an array of length $m$ called a prefix table.

 The main hurdle we must tackle is to store both a prefix table suitable for parameterized matching as well as an encoding of the pattern in only $O(|\SigmaP|+\rho)$ space, while still allowing efficient access to both. It is well-known that any string $P$ can be stored in space proportional to its exact period. In Lemma~\ref{lem:pred-const}, which follows from Lemma~\ref{lem:split}, we show an analogous result for $\pred{P}$.
 See Appendix~\ref{appendix:smallrho} for proofs.

\begin{lemma}\label{lem:split}
For any $j\in [\rho]$ there is a constant $k_j$ such that $\pred{P}[j+k\rho]$ is 0 for $k<k_j$, and $c_j$ for $k\geq k_j$, where $c_j>0$ is a constant that depends on $j$.
\end{lemma}

\begin{lemma}
    \label{lem:pred-const}
    The predecessor string $\pred{P}$ can be stored in $O(\rho)$ space, where $\rho$ is the \pperiod of~$P$. Further, for any $j\in [m]$ we can obtain $\pred{P}[j]$ from this representation in $O(1)$ time.
\end{lemma}

We now explain how to store the parameterized prefix table in only $O(\rho)$ space, in contrast to $\Theta(m)$ space which a standard prefix table would require. The \pperiod $\rho_r$ of $P[0\upto r]$ is, as a function of $r$, non-decreasing in~$r$. This property enables us to run-length encode the prefix table and store it as a doubly linked list with at most $\rho$ elements, hence using only $O(\rho)$ space. Each element corresponds to an interval of prefix lengths with the same \pperiod, and the elements are linked together in increasing order (of the common \pperiod).
This representation does not allow $O(1)$ time random access to the \pperiod of any prefix, however, for our purposes it will suffice to perform sequential access.
To accelerate computation we also store a second linked list of the indices of the first occurrences of each symbol in $P$ in ascending order, i.e. every $j$ such that $\pred{P}[j]=0$. This uses $O(|\SigmaP|)$ space.

There is a crucial second advantage to compressing the prefix table which is that it allows us to upper bound the number of prefixes of $P$ we need to inspect when a mismatch occurs. When a mismatch occurs in our algorithm, we repeatedly shift the pattern until a \pmatch between a text suffix and pattern prefix occurs. Naively it seems that we might have to check many prefixes within the same run.
However, as a consequence of Lemma~\ref{lem:split}
we are assured that if some prefix does not \pmatch, every prefix in the same run  with $\pred{P}[j]\neq 0$ will also mismatch (except possibly the longest). Therefore we can skip inspecting these prefixes.
This can be seen by observing (using Lemma~\ref{lem:split}) that for $j$ such that $\rho_j = \rho_{j+1}$, we have $\pred{P}[j-\rho_j] \in \{0,\,\,\pred{P}[j]\}$.
By keeping pointers into both linked lists, it is straightforward to find the next prefix to check in $O(1)$ time.
Whenever we perform a pattern shift we move at least one of the pointers to the left. Therefore the total number of pattern shifts inspected while processing $T[i]$ is at most $O(|\SigmaP| + \rho)$. As each pointer only moves to the right by at most one when each $T[i]$ arrives, an amortised time complexity of $O(1)$ per character follows. The space usage is $O(|\SigmaP| + \rho)$ as claimed, dominated by the linked lists.

We now briefly discuss how to deamortise our solution by applying Galil's KMP deamortisation argument~\cite{Galil:1981}. The main idea is to restrict the algorithm to shift the pattern at most twice when each text character arrives, giving a constant time algorithm. If we have not finished processing $T[i]$ by this point we accept $T[i+1]$ but place it on the end of a buffer, output `no match' and continue processing $T[i]$. The key property is that the number of text arrivals until the next \pmatch occurs is at least the length of the buffer. As we shift the pattern up to twice during each arrival we always clear the buffer before (or as) the next \pmatch occurs. Further, the size of the buffer is always $O(|\SigmaP| + \rho)$. This follows from the observation above that the number of pattern shifts required to process a single text character is $O(|\SigmaP| + \rho)$.
This concludes the algorithm of Theorem~\ref{thm:kmp-main}. Combining this result with the lower bound result of Appendix~\ref{appendix:space} proves Theorem~\ref{thm:kmp-main}.

\section{The proof of Lemma~\ref{lem:arithmetic}}\label{sec:arithmetic-proof}

    In this section we prove the important Lemma~\ref{lem:arithmetic}. Let \ileft denote an arbitrary position in $T$ where $P$ \pmatches. Let $X$ be the set of positions at which $P$ \pmatches within $T[\ileft \upto (\ileft+3m/2-1)]$. We now prove that there exist disjoint sets $Y$ and $\calA$ with the properties set out in the statement of the lemma.

     Let $\alpha$ be the smallest integer such that all distinct symbols in $P$ occur in the prefix $P[0\upto\alpha]$. We begin by showing that $\rho$, the $\pperiod$ of $P$ is at least $\alpha/|\Sigma|$. From the minimality of $\alpha$, we have that $P[\alpha]$ is the leftmost occurrence of some symbol. By the definition of the $\pperiod$, we have that $P[0 \upto (m-1-\rho)]$ \pmatches $P[\rho\upto m-1]$. Under this shift, $P[\alpha]$ (in $P[\rho\upto m-1]$) is aligned with $P[\alpha-\rho]$ (in $P[0 \upto (m-1-\rho)]$) . Assume that $P[\alpha-\rho]$ is not a leftmost occurrence and let $j$ be the position of the previous occurrence of $P[j]=P[\alpha-\rho]$. As a parameterized match occurs, we have that $P[j]=P[j+\alpha] \neq P[\alpha]$, contradiction. By repeating this argument we have found distinct symbols at positions $\alpha-k\rho$ for all $k>0$. This immediately implies that $\rho\geq \alpha/|\Sigma|$.

     We first deal with two simple cases: $\rho > m/8$ or $\alpha\geq m/4$ (which implies that $\rho\geq m/(4|\Sigma|)$). In these two cases the number of \pmatches is easily upper bounded by $6|\Sigma|$, so all positions can be stored in the set $Y$.

     We therefore continue under the assumption that $\alpha< m/4$ and $\rho< m/8$. As $\rho\geq \alpha/|\Sigma|$, there are at most $(\alpha+1)/(\alpha/|\Sigma|)\leq 2|\Sigma|$ positions from the range $[\ileft \upto \ileft+\alpha]$ at which $P$ can parameterize match $T$. We can store these positions in the set $Y$. Next we will show that the positions from the range $[(\ileft+\alpha+1) \upto (\ileft+3m/2-1)]$ at which $P$ parameterize matches $T$ can be represented by the arithmetic progression~$\calA$.

First we show that $\rho$ is an \emph{exact period} (not \pperiod) of $\pred{P}[\alpha+1\upto m-1]$ (but not necessarily the shortest period). Consider arbitrary positions $P[j]$ and $P[j-\rho]$ where $\alpha<j<m-\rho$. By the definition of the $\pperiod$, we have that $P[\rho\upto m-1]$ \pmatches $P[0 \upto (m-1-\rho)]$ and hence that $\pred{P[\rho\upto m-1]}= \pred{P[0 \upto (m-1-\rho)]}$.
In particular, $\pred{P[\rho\upto m-1]}[j]= \pred{P[0 \upto (m-1-\rho)]}[j]=\pred{P}[j]$,
where the second equality follows because we take the predecessor string of a prefix of $P$.
Also observe that $\pred{P[\rho\upto m-1]}[j]$ either equals $0$ or $\pred{P}[j-\rho]$ by definition. Further, $\pred{P[0 \upto (m-1-\rho)]}[j]= \pred{P}[j] \neq 0$ as $j>\alpha$ and all leftmost occurrences are before $\alpha$.
This implies that $\pred{P[\rho\upto m-1]}[j] \neq 0$, hence, as required, $\pred{P}[j-\rho]=\pred{P[\rho\upto m-1]}[j]=\pred{P[0 \upto (m-1-\rho)]}[j]=\pred{P}[j].$

Recall that $P$ \pmatches $T[\ileft \upto \ileft+m-1]$ so $\pred{P}=\pred{T[\ileft \upto \ileft+m-1]}]$ and hence $\rho$ is an exact period of $\pred{T[\ileft \upto \ileft+m-1]}[\alpha+1\upto m-1]$. Let $j\in\{\alpha+1,\dots,m-2\}$
and observe that by definition, $\pred{T[\ileft \upto \ileft+m-1]}[j] \in \{ 0,\, \pred{T}[\ileft +j]\}$.
However,  $\pred{T[\ileft \upto (\ileft+m-1)]}[j]= \pred{P}[j] >0$ because $j>\alpha$ and all leftmost occurrences are in $P[0 \upto \alpha]$. This implies that $\pred{T[\ileft \upto (\ileft+m-1)]}[j]=\pred{T}[\ileft +j]$. As $j$ was arbitrary, we have that $\pred{T}[(\ileft+\alpha+1) \upto (\ileft+m-1)]=\pred{T[\ileft \upto (\ileft+m-1)]}[\alpha+1\upto m-1]$ and hence $\rho$ is an exact period of $\pred{T}[(\ileft+\alpha+1) \upto (\ileft+m-1)]$.

Let $\iright$ be the rightmost position in $T[\ileft \upto \ileft+3m/2-1]$ where $P$ \pmatches. By the same argument as for $\ileft$, we have that $\rho$ is an exact period of $\pred{T}[(\iright+\alpha+1) \upto (\iright+m-1)]$.

Thus, both $\pred{T}[(\ileft+\alpha+1) \upto (\ileft+m-1)]$ and $\pred{T}[(\iright+\alpha+1) \upto (\iright+m-1)]$ has an exact period of $\rho$. As these two strings overlap by at least $\rho$ characters, we have that $\rho$ is also an exact period of $\pred{T}[\ileft+\alpha+1 \upto \iright+m-1]$.

Let $i\in\{(\ileft+\alpha+1),\dots,\iright-1\}$ be arbitrary such that $P$ \pmatches $T[i \upto (i+m-1)]$. We now prove that if $i+\rho<\iright$ then $P$ \pmatches $T[i+\rho \upto (i+\rho+m-1)]$. As \pmatches must be at least $\rho$ characters apart this is sufficient to conclude that all remaining matches form an arithmetic progression with common difference $\rho$.

As $\rho$ is an exact period of $\pred{T}[(\ileft+\alpha+1) \upto (\iright+m-1)]$, we have that $\pred{T}[i \upto (i+m-1)]=\pred{T}[i+\rho \upto (i+\rho+m-1)]$. By definition, this implies that $\pred{T[i \upto (i+m-1)]}=\pred{T[i+\rho \upto (i+\rho+m-1)]}$ and hence a \pmatch also occurs at $i+\rho$. This concludes the proof of Lemma~\ref{lem:arithmetic}.

\section{Acknowledgements}

The work described in this paper was supported by EPSRC. The authors would like to thanks Rapha\"el Clifford for many helpful and encouraging discussions.

\printbibliography

\newpage

\appendix

\markboth{APPENDIX}{APPENDIX}

\section{Space lower bounds}\label{appendix:space}

To complete the picture we give nearly matching space lower bounds which show that our
solutions are optimal to within log factors.  The proof is by a
communication complexity argument. In
essence one can show that in the randomised case Alice is able to
transmit any string of length $\Theta(|\SigmaP|)$ bits to Bob
using a solution to the matching problem by selecting a suitable pattern and streaming text.  Similarly in the deterministic case (see below) one can show
that she can send $\Theta(|\SigmaP|+\rho)$ bits.

\begin{proof}[\prooftheorem{thm:space}]
Consider first a pattern where all symbols are distinct, e.g.\@ $P=\texttt{123456}$.  Now let us assume Alice would like to send a bit-string to Bob. She can encode the bit-string as an instance of the parameterized matching problem in the following way. As an example, assume the bit-string is \texttt{01011}. She first creates the first half of a text stream \texttt{aBcDE} where we choose capitals to correspond to \texttt{1} and lower case symbols to correspond to \texttt{0} from the original bit-string.  She starts the matching algorithm and runs it until the pattern and the first half of the text have been processed and then sends a snapshot of the memory to Bob.  Bob then continues with the second half of the text which is fixed to be the sorted lower case symbols, in this case \texttt{abcde}.  Where Bob finds a parameterized match he outputs a \texttt{1} and where he does not, he outputs a \texttt{0}.  Thus Alice's bit-string is reproduced by Bob. In general, if we restrict the alphabet size of the pattern to be $|\SigmaP|$ then Alice can similarly encode a bit-string of length $|\SigmaP|-1$, and successfully transmit it to Bob, giving us an $\Omega(|\SigmaP|)$ bit lower bound on the space requirements of any streaming algorithm.
\end{proof}

If randomisation is not allowed, the lower bound increases to $\Omega(|\SigmaP| + \rho)$ bits of space. Here $\rho$ is the parameterized period of the pattern. This bound follows by a similar argument by devising a one-to-one encoding of bit-strings of length $\Theta(\rho)$ into $P[0 \ldots \rho -1]$. The key difference is that with a deterministic algorithm, Bob can enumerate all possible $m$-length texts to recover Alice's bit-string from~$P$.

\section{Correctness proof of the main algorithm}\label{appendix:correctness}

\begin{proof}[\prooflemma{lem:correctness}]
    Coupled with the discussion in Section~\ref{sec:overview}, the time and space complexity almost follow immediately from the description. It only remains to show that, at any time, $|\calB| \leq  |\SigmaP|$. First observe that any symbol $\sigma \in \Sigma_T$ is only inserted into $\calB$ when $\pred{T}[i]>m_0>\delta$ which can only happen at most once in every $\delta = |\SigmaP| \log{m}$ arriving symbols. Further we remove one element every $s \leq \lceil \log m \rceil$ arrivals and in particular remove the $\sigma$ occurrence after at most $|\calB| \lceil \log m \rceil$ arrivals. As $\calB$ is initially empty, by induction it follows that no symbol occurs more than once in $\calB$.

    For correctness, it remains to show that we correctly obtain the positions of $\fpinner{\ell}(i')$ from $D_\ell$. It follows from the description that all positions of $\fpinner{\ell}(i')$ correspond to elements inserted into $D_\ell$ at some point. However we need to prove that these elements are present in $D_\ell$ while $\fpinner{\ell}(i')$ is calculated. Any element inserted into $\calB$ during $T[i' \upto (i'+m_{\ell}-1)]$ has cleared the buffer by the end of interval B (which has length $\delta$) by the argument above. Therefore any relevant element has been inserted into $D_\ell$ by the start of interval C, during which we calculate $\fpinner{\ell}(i')$. Any element inserted into $D_\ell$ is at least $m_{\ell-1}$ characters from its predecessor. Therefore, summing over all symbols in the alphabet, there are at most $4|\Sigma_P|$ positions in $T[i' \upto (i'+2m_{\ell}-1)]$ which are inserted into $D_\ell$. As $D_\ell$ is a FIFO queue of size $12|\Sigma|$, the relevant elements are still present after interval C.

    As commented earlier, potential matches in $M_\ell$ are separated by more than $3\delta$ arrivals because $P_{\ell-1}$ has $\pperiod$ more than $3\delta$. They are processed within $3\delta$ arrivals so $M_\ell$ does not overflow. This completes the correctness.
\end{proof}

\section{Proof of Theorem~\ref{thm:filter} (general alphabets)} \label{appendix:filter}

Let $\Sigma_T$ denote the text alphabet.
In order to handle general alphabets we perform two reductions in sequence on each arriving text symbol (and on $P$ during preprocessing). The first reduces $\Sigma_P$ and $\Sigma_T$ to each contain only symbols from $\Pi$ and one additional variable symbol (which is different for $P$ and $T$). A suitable such reduction is given in~\cite{AFM:1994} (Lemma 2.2). The reduction is presented for the offline version but immediately generalises by using the constant time exact matching algorithm of Breslauer and Galil~\cite{BG:2011}.

We now define $\SigmaP'$ to be the pattern alphabet after the first reduction (and $\SigmaT'$ respectively). Note that $|\SigmaP'|=|\SigmaT'|=|\Pi|+1$ and all pattern symbols are variables. However we have no guarantee on the bit representations of the alphabet symbols. Let $T'$ and $P'$ denote the text and pattern after the first reduction. The second reduction now maps each $T'[i]$ into the range $\{0,\ldots,|\SigmaP'|\}$ as it arrives. The equivalent reduction for the pattern is a simplification which can be performed in preprocessing.

Let the strings $\Sm$ and $\SmFilt$ denote the last $m$ characters of the unfiltered (post first reduction) and filtered (post second reduction) stream, respectively. Let $\SigmaLast\subseteq \SigmaT'$ denote the up to $|\SigmaP'|+1$ last distinct symbols in $\Sm$, hence $|\SigmaLast|$ is never more than $|\SigmaP'|+1$. Let $\calT$ be a dynamic dictionary on $\SigmaLast$ such that a symbol in $\SigmaT'$ can be looked up, deleted and added in $O(\sqrt{\log{|\SigmaP'|}/\log{\log{|\SigmaP'|}}})$ time~\cite{AT:2000}. Every symbol that arrives in the stream is associated with its ``arrival time'', which is an integer that increases by one for every new symbol arriving in the stream. Let $\calL$ be an ordered list of the symbols in $\SigmaLast$ (together with their most recent arrival time) such that $\calL$ is ordered according to the most recent arrival time. For example,
    \begin{equation}
        \label{eq:list}
        \calL = (\texttt{d}, 25),\, (\texttt{b}, 33),\, (\texttt{g}, 58),\, (\texttt{e}, 102)
    \end{equation}
    means that the symbols \texttt{b}, \texttt{d}, \texttt{e} and \texttt{g} are the last four distinct symbols that appear in $\Sm$ (for this example, $|\SigmaP'|+1\geq 4$), where the last \texttt{e} arrived at time~102, the last \texttt{g} arrived at time~58, and so on.

    By using appropriate pointers between elements of the hash table~$\calT$ and elements of $\calL$ (which could be implemented as a linked list), we can maintain $\calT$ and $\calL$ in $O(1)$ time per arriving symbol. To see this, take the example in Equation~(\ref{eq:list}) and consider the arrival of a new symbol $x$ at time~103 (following the last symbol \texttt{e}). First we look up $x$ in $\calT$ and if $x$ already exists in $\SigmaLast$, move it to the right end of $\calL$ by deleting and inserting where needed and update the element to $(x,103)$. Also check that the leftmost element of $\calL$ is not a symbol that has been pushed outside of $\Sm$ when $x$ arrived. We use its arrival time to determine this and remove the last element accordingly. If the arriving symbol $x$ does not already exist in $\SigmaLast$, then we add $(x,103)$ to the right end of $\calL$. To ensure that $\calL$ does not contain more than $|\SigmaP'|+1$ elements, we remove the leftmost element of $\calL$ if necessary. We also remove the leftmost symbol if it has been pushed outside of $\Sm$. The hash table $\calT$ is of course updated accordingly as well.

    Let $\SigmaFilt=\{0,\dots,|\SigmaP'|\}$ denote the symbols outputted by the filter. We augment the elements of $\calL$ to maintain a mapping $\calM$ from the symbols in $\SigmaLast$ to distinct symbols in $\SigmaFilt$ as follows. Whenever a new symbol is added to $\SigmaLast$, map it to an unused symbol in $\SigmaFilt$. If no such symbol exists, then use the symbol that is associated with the symbol of $\SigmaLast$ that is to be removed from $\SigmaLast$ (note that $|\SigmaLast|\leq |\SigmaFilt|$). The mapping $\calM$ specifies the filtered stream: when a symbol $x$ arrives, the filter outputs $\calM(x)$. Finding $\calM(x)$ and updating $\calT$ is done in $O(1)$ time per arriving character, and both the tree $\calT$ and the list $\calL$ can be stored in $O(|\SigmaP'|)$ space.

    It remains to show that the filtered stream does not induce any false matches or miss a potential match. Suppose first that the number of distinct symbols in $\Sm$ is $|\SigmaP'|$ or fewer. That is, $\SigmaLast$ contains all distinct symbols in $\Sm$. Every symbol $x$ in $\Sm$ has been replaced by a unique symbol in $\SigmaFilt$ and the construction of the filter ensures that the mapping is one-to-one. Thus, $\pred{\SmFilt}=\pred{\Sm}$. Suppose second that the number of distinct symbols in $\Sm$ is $|\SigmaP'|+1$ or more. That is, $|\SigmaLast|=|\SigmaP'|+1$ and therefore $\SmFilt$ contains $|\SigmaP'|+1$ distinct symbols. Thus, $\pred{\SmFilt}$ cannot equal $\pred{P'}$. The claimed result then follows from Theorem~\ref{thm:main}.

\section{Proofs omitted from Section~\ref{sec:smallrho}} \label{appendix:smallrho}

\begin{proof}[\prooflemma{lem:split}]
    Let $\rho$ be the \pperiod of $P$. We prove the lemma by contradiction. Suppose, for some $j$ and $k$, that $i=j+k\rho$ is a position such that $\pred{P}[i]=c\geq 1$ and $\pred{P}[i+\rho]=c'\neq c$. Consider Figure~\ref{fig:pattern-rho} for a concrete example, where $\rho=5$, $i=12$, $\pred{P}[12]=c=4$ and $\pred{P}[12+5]=c'=3$.
    \begin{figure}[t]
        \centering
        \includegraphics[scale=0.90]{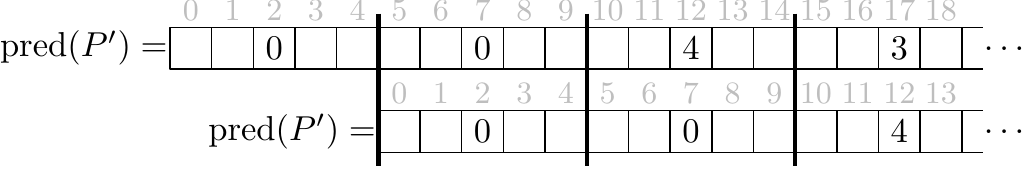}
        \caption{\label{fig:pattern-rho} An example demonstrating the structure of $\pred{P'}$ used in the proof of Lemma~\ref{lem:split}.}
    \end{figure}
    Since $\rho$ is a \pperiod of $P$, we have that
    \begin{equation*}
        \pred{P[\rho\upto m-1]} = \pred{P[0\upto (m-1-\rho)]} \,.
    \end{equation*}
    Consider the alignment of positions $i+\rho$ and $i$ (positions 17 and~12 in Figure~\ref{fig:pattern-rho}). We have that $\pred{P[\rho\upto m-1]}[i]$ is either $c'$ or~0. In either case, it is certainly not $\pred{P[0\upto m-1-\rho]}[i]$ which is $c$. Thus, $\rho$ cannot be a \pperiod of $P$.
\end{proof}

\begin{proof}[\prooflemma{lem:pred-const}]
    By Lemma~\ref{lem:split} we can encode $\pred{P}$ by storing the two values $k_j$ and $c_j$ for each $j\in [\rho]$. This takes $O(\rho)$ space. The value $\pred{P}[i]$ is 0 if $i<k_{(i\bmod \rho)}$, otherwise it is $c_{(i\bmod \rho)}$.

\end{proof}

\end{document}